%% file: main.tex
\documentclass{article}
\usepackage{ijcai18}
\usepackage{times}
\usepackage{xcolor}
\usepackage{soul}
\usepackage[utf8]{inputenc}
\usepackage[small]{caption}

\usepackage{hyperref}
\usepackage{url}
\usepackage{makecell}

\usepackage{booktabs}
\usepackage{makecell}
\usepackage{multirow}
\usepackage[utf8]{inputenc} % allow utf-8 input
\usepackage[T1]{fontenc}    % use 8-bit T1 fonts
\usepackage{balance} % For balanced columns on the last page
\usepackage{amssymb}
\usepackage{array}
\usepackage{multirow}
\usepackage[linesnumbered,ruled]{algorithm2e}
\usepackage{algpseudocode}
\usepackage{amsmath,amssymb,mathtools}
\usepackage{cuted}
\usepackage{amsthm,bm}
\usepackage{afterpage}
\usepackage{bbm}
\usepackage{color}
\usepackage{algorithm2e}
\usepackage{url}

\newcommand{\bfx}{x}
\newcommand{\bfxA}{x_A}

\SetCommentSty{mycommfont}

\usepackage{array}
\usepackage{multirow}
\usepackage[linesnumbered,ruled]{algorithm2e}
\usepackage{amsmath,amssymb,mathtools}
\usepackage{cuted}
\usepackage{amsthm}
\usepackage{afterpage}
\usepackage{color}

\newcommand{\reffig}[1]{Figure~\ref{fig:#1}}

\newcommand{\lblfig}[1]{\label{fig:#1}}
\newcommand{\lblsec}[1]{\label{sec:#1}}
\newcommand{\lbleq}[1]{\label{eq:#1}}

\newcommand{\ignorethis}[1]{}
\newcommand{\norm}[1]{\lVert#1\rVert}

\usepackage{subcaption}
\usepackage{array}
\usepackage{multirow}
\usepackage[linesnumbered,ruled]{algorithm2e}
\usepackage{algpseudocode}
\usepackage{amsmath,amssymb,mathtools}
\usepackage{cuted}
\usepackage{amsthm}
\usepackage{afterpage}
\usepackage{color}
\usepackage{algorithm2e}
\usepackage[export]{adjustbox}% http://ctan.org/pkg/adjustbox
\usepackage{times}
\newif\ifsubmit
\submitfalse
\ifsubmit
\newcommand{\bo}[1]{}
\newcommand{\junyanz}[1]{}
\newcommand{\chaowei}[1]{}
\newcommand{\warren}[1]{}
\newcommand{\rachel}[1]{}
\newcommand{\richard}[1]{}
\else
\newcommand{\bo}[1]{\textcolor{blue}{Bo: #1}}

\newcommand{\junyanz}[1]{\textcolor{red}{JY: #1}}
\newcommand{\rachel}[1]{\textcolor{red}{Rachel: #1}}
\newcommand{\chaowei}[1]{\textcolor{magenta}{chao: #1}}

\newcommand{\warren}[1]{\textcolor{cyan}{(Warren: #1)}}
\newcommand{\richard}[1]{\textcolor{magenta}{Richard: #1}}

\fi

\title{Generating Adversarial Examples with Adversarial Networks}
\author{
Chaowei Xiao$^{1}$ \thanks{This work was performed when Chaowei Xiao was at JD.COM and University of California, Berkeley}, 
Bo Li$^2$, 
Jun-Yan Zhu$^{2,3}$, 
Warren He$^2$,
Mingyan Liu$^1$ \and 
Dawn Song$^2$
\\ 
$^1$University of Michigan, Ann Arbor \\
$^2$University of California, Berkeley \\
$^3$Massachusetts Institute of Technology\\
}
\begin{document}

\maketitle

\input{def}
% \vspace{-9cm}
\begin{abstract}
Deep neural networks (DNNs) have been found to be vulnerable to adversarial examples resulting from adding small-magnitude perturbations to inputs. Such adversarial examples can mislead DNNs to produce adversary-selected results.
Different attack strategies have been proposed to generate adversarial examples, but how to produce them with high perceptual quality and more efficiently requires more research efforts. 
In this paper, we propose AdvGAN to generate adversarial examples with generative adversarial networks (GANs), which can learn and approximate the distribution of original instances. 
For AdvGAN, once the generator is trained, it can generate perturbations efficiently for any instance, so as to potentially accelerate adversarial training as defenses.  
We apply AdvGAN in both \white and black-box attack settings. In \white attacks, there is no need to access the original target model after the generator is trained, in contrast to traditional white-box attacks. In black-box attacks, we dynamically train a distilled model for the black-box model and optimize the generator accordingly.
Adversarial examples generated by AdvGAN on different target models have high attack success rate under state-of-the-art defenses compared to other attacks. Our attack  has placed the first with 92.76\% accuracy on a public MNIST black-box attack challenge.\footnote{https://github.com/MadryLab/mnist\_challenge}
\end{abstract} 

% \vspace{-9cm}
% \vspace{-0.2cm}
\section{Introduction}
\vspace{-0.1cm}
Deep Neural Networks (DNNs) have achieved great successes in a variety of applications.
However, recent work has demonstrated that DNNs are vulnerable to adversarial perturbations \cite{szegedy2014intriguing,goodfellow2014explaining,hu2017generating}. 
An adversary can add small-magnitude perturbations to inputs and generate adversarial examples to mislead DNNs.
Such maliciously perturbed instances can cause the learning system to misclassify them into either a maliciously-chosen target class (in a targeted attack) or classes that are different from %with 
the ground truth (in an untargeted attack).
Different algorithms have been proposed for generating such adversarial examples, such as the fast gradient sign method (FGSM) \cite{goodfellow2014explaining} and optimization-based methods (Opt.) \cite{carlini2017towards,liu2016delving,xiao2018spatially,evtimov2017robust}.

Most of the the current attack algorithms \cite{carlini2017towards,liu2016delving} rely on optimization schemes with simple pixel space metrics, such as $L_\infty$ distance from a benign image, to encourage visual realism.
To generate more perceptually realistic adversarial examples efficiently,  
in this paper, we propose to train (i) a feed-forward network that generate perturbations to create diverse adversarial examples and (ii) a discriminator network to ensure that the generated examples are realistic. We apply generative adversarial networks (GANs) \cite{goodfellow2014generative} to produce adversarial examples
in both the \white and black-box settings. 
As conditional GANs are capable of producing high-quality images \cite{isola2017image}, we apply a similar paradigm to produce perceptually realistic adversarial instances. 
We name our method AdvGAN.

Note that in the previous white-box attacks, such as FGSM and optimization methods, the adversary needs to have white-box access to the architecture and parameters of the model all the time. However, by deploying AdvGAN, once the feed-forward network is trained, it can instantly produce adversarial perturbations for any input instances without requiring access to the model itself anymore. We name this attack setting \emph{\white}.

To evaluate the effectiveness of our attack strategy AdvGAN , we first generate adversarial instances based on AdvGAN and other attack strategies on different target models. We then apply the state-of-the-art defenses to defend against these generated adversarial examples \cite{goodfellow2014explaining,madry_towards_2017}. We evaluate these attack strategies in both \white and black-box settings.
We show that adversarial examples generated by AdvGAN can achieve a high attack success rate, potentially due to the fact that these adversarial instances appear closer to real instances compared to other recent attack strategies. 

Our contributions are listed as follows.
\begin{itemize}
    \item Different from the previous optimization-based methods, we train a conditional adversarial network to directly produce adversarial examples, which not only results in perceptually realistic examples that achieve state-of-the-art attack success rate against different target models, but also the generation process is more efficient.
    \item We show that AdvGAN can attack black-box models by training a distilled model. We propose to dynamically train the distilled model with query information and achieve high black-box attack success rate and targeted black-box attack, which is difficult to achieve for transferability-based black-box attacks. 
    \item 
    We use the state-of-the-art defense methods to defend against adversarial examples and show that AdvGAN 
    achieves much higher attack success rate under current defenses.
    \item We apply AdvGAN on \citeauthor{madry_towards_2017}'s MNIST challenge (\citeyear{madry_towards_2017}) and achieve 88.93\% accuracy on the published robust model in the semi-whitebox setting and 92.76\% in the black-box setting, which wins the top position in the challenge.
\end{itemize}

% \vspace{-0.2cm}
\section{Related Work}
\vspace{-0.1cm}
Here we review recent work on adversarial examples and generative adversarial networks.

\textbf{Adversarial Examples}
\lblsec{related_adv}
A number of attack strategies to generate adversarial examples have been proposed in the white-box setting, where the adversary has full access to the classifier \cite{szegedy2014intriguing,goodfellow2014explaining,carlini2017towards,xiao2018spatially,hu2017generating}.
% \textbf{Fast Gradient Method.}
\citeauthor{goodfellow2014explaining} propose the fast gradient sign method (FGSM), which applies a first-order approximation of the loss function to construct adversarial samples. 
Formally, given an instance $\bfx$, an adversary generates adversarial example $\bfxA = \bfx + \eta$ with $L_{\infty}$ constraints  
% \warren{other method names are set in italic, not capitalized}
in the untargeted attack setting as
$\eta= \epsilon \cdot \text{sign} (\nabla_{\bfx}\ell_f(\bfx, y))$,
where $\ell_f(\cdot)$ is the cross-entropy loss used to train the neural network $f$, and
$y$ 
represents the ground truth of $\bfx$.
Optimization based methods (Opt)
have also been proposed to optimize adversarial perturbation for targeted attacks while satisfying certain constraints \cite{carlini2017towards,liu2016delving}. 
Its goal is to minimize the objective function as $ ||\eta|| + \lambda \ell_f(\bfxA, y)$, 
where $||\cdot||$ is an appropriately chosen norm function. 
However, the optimization process is slow and can only optimize perturbation for one specific instance each time. 
 In contrast, our method uses feed-forward network to generate an adversarial image, rather than an optimization procedure.  Our method achieves higher attack success rate against different defenses and performs much faster than the current attack algorithms.

Independently from our work, 
feed-forward networks have been applied to generate adversarial perturbation \cite{baluja2017adversarial}.
However, \citeauthor{baluja2017adversarial} combine the re-ranking loss and  an $L_2$ norm loss, aiming to constrain the generated adversarial instance to be close to the original one in terms of $L_2$; while we apply a deep neural network as a discriminator to help distinguish the instance with other real images to encourage the perceptual quality of the generated adversarial examples .
\citeauthor{hu2017generating}\cite{hu2017generating} also proposed to use GAN to generate adversarial examples. However, they aim to generate adversarial examples for malware while our work focus on generating perceptual realistic adversarial examples for image.

\textbf{Black-box Attacks} 
Current learning systems usually do not allow white-box accesses against the model for security reasons. Therefore, there is a great need for black-box attacks analysis.

Most of the black-box attack strategies are based on the transferability phenomenon \cite{papernot2016practical}, 
where an adversary can train a local model first and generate adversarial examples against it, hoping the same adversarial examples will also be able to attack the other models. 
% \textbf{Query based attacks.}
Many learning systems allow query accesses to the model. 
However, there is little work that can leverage query-based access to target models to construct adversarial samples and move beyond transferability. \citeauthor{hu2017generating} proposed to leverage GANs to construct evasion instance for malware.    
\citeauthor{papernot2016practical}  proposed to train a local substitute model with queries to the target model to generate adversarial samples, but this strategy still relies on transferability.
In contrast, we show that the proposed AdvGAN can perform black-box attacks without 
depending on transferability. 

\textbf{Generative Adversarial Networks (GANs)}  \citeauthor{goodfellow2014generative}  have achieved visually appealing results in both image generation and manipulation \cite{zhu2016generative} settings. Recently, image-to-image conditional GANs have further improved the quality of synthesis results \cite{isola2017image}. We adopt a similar adversarial loss and image-to-image network architecture to learn the mapping from an original image to a perturbed output such that the perturbed image cannot be distinguished from real images in the original class. Different from prior work, we aim to produce output results that are not only visually realistic but also able to mislead target learning models. 
\section{Generating Adversarial Examples with Adversarial Networks}
\lblsec{adv_gan}
\subsection{Problem Definition}
Let $\mathcal{X} \subseteq \mathcal{R}^n$ be the feature space, with $n$ the number of features.
Suppose that $(\bfx_i,y_i)$ is the $i$th instance within the training set, which is comprised of feature vectors $\bfx_i \in \mathcal{X}$, generated according to some unknown distribution $\bfx_i \sim \mathcal{P}_{\text{data}}$, and $y_i \in \mathcal{Y}$ the corresponding true class labels.
The learning system aims to learn a classifier $f: \mathcal{X} \rightarrow \mathcal{Y}$ from the domain $\mathcal{X}$ to the set of classification outputs $\mathcal{Y}$,
%(e.g., $\mathcal{Y} \in \{0,1\}$ for binary classification). \warren{why binary classification as the example?} 
where $|\mathcal{Y}|$ denotes the number of possible classification outputs. 
Given an instance $\bfx$, the goal of an adversary is to generate adversarial example $\bfx_{A}$, which is classified as $f(\bfx_{A}) \ne y$ (untargeted attack), where $y$ denotes the true label; or $f(\bfx_{A}) = t$ (targeted attack) where $t$ is the target class. 
$\bfx_{A}$ should also be close to the original instance $\bfx$ in terms of $L_2$ or other distance metric.

\begin{figure}[t!]
 \centering
 \includegraphics[width=0.45\textwidth]{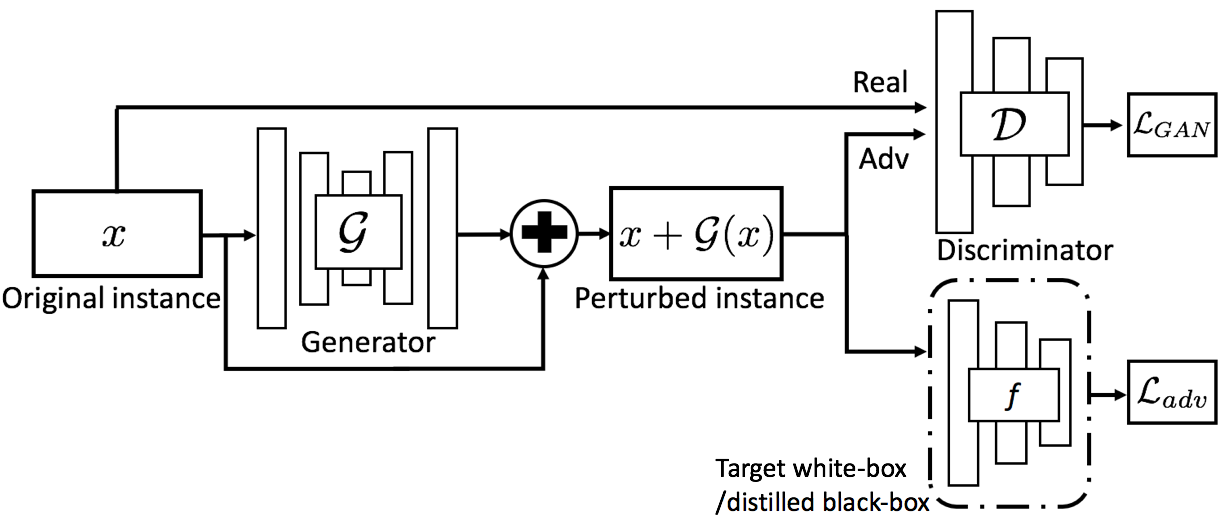}
 \caption{Overview of AdvGAN\vspace{-0.2cm}}
 \lblfig{framework}
\end{figure}
\subsection{AdvGAN Framework}
\label{s:whitebox}
\reffig{framework} illustrates the overall architecture of AdvGAN, which mainly consists of three parts: a generator $\mathcal{G}$, a discriminator $\mathcal{D}$, and the target neural network  $f$.
Here the generator $\mathcal{G}$ takes the original instance $x$ as its input and generates a perturbation $\mathcal{G}(x)$.  Then $x+\mathcal{G}(x)$ will be sent to the discriminator $\mathcal{D}$, which is used to distinguish the generated data and the original instance $x$. The goal of $\mathcal{D}$ is to encourage that the generated instance is indistinguishable with the data from its original class. To fulfill the goal of fooling a learning model, we first perform the white-box attack, where the target model is $f$ in this case. $f$ takes $x+\mathcal{G}(x)$ as its input and outputs its loss $\mathcal{L}_{adv}$, which represents the distance between the prediction and the target class $t$ (targeted attack), or the opposite of the distance between the prediction and the ground truth class (untargeted attack).

The adversarial loss \cite{goodfellow2014generative} can be written as:\footnote{For simplicity, we denote the $\mathbb{E}_x \equiv \mathbb{E}_{x \sim \mathcal{P}_{\text{data}}}$}
\vspace{-0.1cm}
\begin{equation}
\vspace{-0.1cm}
\mathcal{L}_{\text{GAN}} = \mathbb{E}_{x} \log \mathcal{D}(x) +  \mathbb{E}_{x} \log(1-\mathcal{D}(x+\mathcal{G}(x))).
\end{equation}
Here, the discriminator $\mathcal{D}$ aims to distinguish the perturbed data $x+\mathcal{G}(x)$ from the original data $x$.
% \footnote{Note that we only use the generator to produce the perturbation $\mathcal{G}(x)$.}
Note that the real data is sampled from the true class, so as to encourage that the generated instances are close to data from the original class.

The loss for fooling the target model $f$ in a targeted attack is:
\vspace{-0.1cm}
\begin{equation}
\mathcal{L}_{\text{adv}}^f = \mathbb{E}_{x}\ell_f (x+\mathcal{G}(x),t),
\end{equation}
where $t$ is the target class and $\ell_f$ denotes the loss function (e.g., cross-entropy loss) used to train the original model $f$. The $\mathcal{L}_{adv}^f$ loss encourages the perturbed image to be misclassified as target class $t$.
Here we can also perform the untargeted attack by maximizing the distance between the prediction and the ground truth, but we will focus on the targeted attack in the rest of the paper.
% \dawn{we should talk about untargated attack as well. and say we omit the details since it's simple.}\bo{done}
%\chaowei{need to discuss}

To bound the magnitude of the perturbation, which is a common practice in prior work \cite{carlini2017towards,liu2016delving}, we add a soft hinge loss on the $L_2$ norm as
\vspace{-0.1cm}
\begin{equation}
\label{eq:hingeloss}
\mathcal{L}_{\text{hinge}}=\mathbb{E}_{x} \max(0, \norm{\mathcal{G}(x)}_2-c),
\end{equation}
where $c$ denotes a user-specified bound. This can also stabilize the GAN's training, as shown in \citeauthor{isola2017image}\ (\citeyear{isola2017image}). 
Finally, our full objective can be expressed as
\vspace{-0.1cm}
\begin{equation}\lbleq{full_loss}
\mathcal{L} = \mathcal{L}_{\text{adv}}^f +\alpha \mathcal{L}_{\text{GAN}}+ \beta \mathcal{L}_{\text{hinge}},
\end{equation}
where $\alpha$ and $\beta$ control the relative importance of each objective.
Note that $\mathcal{L}_{\text{GAN}}$ here is used to encourage the perturbed data to appear similar to the original data $x$, while $\mathcal{L}^f_{\text{adv}}$ is leveraged to generate adversarial examples, optimizing for the high attack success rate. 
We obtain our $\mathcal{G}$ and $\mathcal{D}$ by solving the minmax game 
 $\arg\min_{\mathcal{G}}\max_{\mathcal{D}} \mathcal{L}$. Once $\mathcal{G}$ is trained on the training data and the target model, it can produce perturbations for any input instance to perform a \white attack.

\vspace{-0.2cm}
\subsection{Black-box Attacks with Adversarial Networks}
\lblsec{blackbox}
\paragraph{Static Distillation}
For black-box attack, we assume adversaries have no prior knowledge of training data or the model itself. In our experiments in Section~\ref{s:exp}, we randomly draw data that is \emph{disjoint} from the training data of the black-box model to distill %distillate 
it, since we assume the adversaries have no prior knowledge about the training data or the model. 
To achieve black-box attacks, we first build a distilled network $f$ based on the output of the black-box model $b$ \cite{hinton2015distilling}. Once we obtain the distilled network $f$, we carry out the same attack strategy as described in the white-box setting (see Equation~\eqref{eq:full_loss}). Here, we minimize the following network distillation objective:
\vspace{-0.1cm}
\begin{equation}
\label{lf}
\arg\min_f \mathbb{E}_{x} \; \mathcal{H}(f(x), b(x)),
\vspace{-0.2cm}
\end{equation}

where $f(x)$ and $b(x)$ denote the output from the distilled model and black-box model respectively for the given training image $x$, and $\mathcal{H}$ denotes the commonly used cross-entropy loss.
By optimizing the objective over all the training images, we can obtain a model $f$ which behaves very close to the black-box model $b$. We then carry out the attack on the distilled network.

Note that unlike training the discriminator $\mathcal{D}$, where we only use the real data from the original class to encourage that the generated instance is close to its original class, here we train the distilled model with data from all classes.
\vspace{-0.4cm}
\paragraph{Dynamic Distillation}
Only training the distilled model with all the pristine training data is not enough, since it is unclear how close the black-box and distilled model perform on the generated adversarial examples, which have not appeared in the training set before. 
Here we propose an \emph{alternative minimization} approach to dynamically make queries and train the distilled model $f$ and our generator $\mathcal{G}$ jointly. 
We perform the following two steps in each iteration. During iteration $i$:

\begin{enumerate}
\item 
{\bf Update $\mathcal{G}_i$  given a fixed network $f_{i-1}$}: We follow the white-box setting (see Equation \ref{eq:full_loss}) and train the generator and discriminator based on a previously distilled model $f_{i-1}$. We initialize the weights $\mathcal{G}_i$ as $\mathcal{G}_{i-1}$.
$\mathcal{G}_{i}, D_{i} = \arg\min_{\mathcal{G}}\max_{\mathcal{D}}
\mathcal{L}_{\text{adv}}^{f_{i-1}} +\alpha \mathcal{L}_{\text{GAN}}+ \beta \mathcal{L}_{\text{hinge}}$
% \end{equation}

\item
{\bf Update $f_i$ given a fixed generator $\mathcal{G}_i$}: First, we use $f_{i-1}$ to initialize $f_i$.
Then, given the generated adversarial examples $x+\mathcal{G}_i(x)$ from $\mathcal{G}_i$, the distilled model $f_i$ will be updated based on the set of new query results for the generated adversarial examples against the black-box model, as well as the original training images.

$f_i = \arg\min_f \mathbb{E}_{x} \mathcal{H}(f(x), b(x)) + \mathbb{E}_{x} \mathcal{H}(f(x+\mathcal{G}_i(x)), b(x+\mathcal{G}_i(x)))$,
where we use both the original images $x$ and the newly generated adversarial examples $x+ s\mathcal{G}_i(x)$ to update $f$.

\end{enumerate}

In the experiment section, we compare the performance of both the static and dynamic distillation approaches and observe that simultaneously updating $\mathcal{G}$ and $f$ produces higher attack performance. See Table~\ref{t-performance} for more details.

\section{Experimental Results}
\label{s:exp}
In this section, we first evaluate AdvGAN for both \white and black-box settings on MNIST \cite{lecun1998mnist} and CIFAR-10 \cite{krizhevsky2014cifar}. We also perform a \white attack on the ImageNet dataset \cite{deng2009imagenet}. We then apply AdvGAN to generate adversarial examples on different target models and test the attack success rate for them under the state-of-the-art defenses and show that our method can achieve higher attack success rates compared to other existing attack strategies.
We generate all adversarial examples for different attack methods under an $L_\infty$ bound of 0.3 on MNIST and 8 on CIFAR-10, for a fair comparison. 
In general, as shown in Table~\ref{t-data-compare}, AdvGAN has several advantages over other white-box and black-box attacks. For instance, regarding computation efficiency, AdvGAN performs much faster than others even including the efficient FGSM, although AdvGAN needs extra training time to train the generator.
All these strategies can perform targeted attack except transferability based attack, although the ensemble strategy can help to improve.
Besides, FGSM and optimization methods can only perform white-box attack, while AdvGAN is able to attack in \white setting.

\begin{table}[t]
\begin{small}
    \centering
     
    \begin{tabular}{ccccc}
        \toprule
         &FGSM & Opt. & Trans.  & AdvGAN\\ \hline 
        \hline
        Run time  & 0.06s &$>$3h &- & <0.01s\\ 
        Targeted Attack  &  \checkmark & \checkmark & Ens. & \checkmark \\ 
        Black-box Attack & & & \checkmark& \checkmark\\
        \bottomrule
    \end{tabular}
  \caption{Comparison with the state-of-the-art attack methods. Run time is measured for generating 1,000 adversarial instances during test time. Opt. represents the optimization based method, and Trans. denotes black-box attacks based on transferability.}
      \label{t-data-compare} 
     \vspace{-0.2cm}

    \end{small}
\end{table}

\noindent{\bf Implementation Details} We adopt similar architectures for generator and discriminator with image-to-image translation literature \cite{isola2017image,zhu2017unpaired}. 
We apply the loss in \citeauthor{carlini2017towards} (\citeyear{carlini2017towards}) as our loss $\mathcal{L}_{\mathit{adv}}^f = \max (\max_{i\neq t} f(\bfxA)_i - f(\bfxA)_t,\kappa)$, where $t$ is the target class, and 
$f$ represents the target network in the \white setting and the distilled model in the black-box setting. 
We set the confidence $\kappa=0$ for both Opt.\ and AdvGAN. 
We use 
%use Adam as our solver \cite{kingma2014adam}, with
a batch size of 128 and a learning rate of 0.001.
For GANs training,  we use the least squares objective proposed by LSGAN \cite{mao2016least}, as it has been shown to produce better results with more stable training.

\begin{table}[t]
\begin{small}
    \centering

    \addtolength{\tabcolsep}{-6pt}
    \begin{tabular}{c|c c c | c c }
        \toprule
        & \multicolumn{3}{ c |}{MNIST(\%)} &  \multicolumn{2}{ c }{CIFAR-10(\%)} \\
       Model & A  & B &C & ResNet & Wide ResNet \\ \hline 
        \hline
    Accuracy (p) & 99.0\;\;\; & 99.2\;\;\; & 99.1  & 92.4 & 95.0\\ \hline \hline
    Attack Success Rate (w) & 97.9\;\;\; & 97.1\;\;\; &  98.3 & 94.7 & 99.3  \\ \hline
    Attack Success Rate (b-D) & 93.4\;\;\; & 90.1\;\;\;&  94.0 & 78.5& 81.8 \\ \hline
    Attack Success Rate (b-S) & 30.7\;\;\; & 66.6\;\;\; &  87.3 & 10.3& 13.3 \\ 

        \bottomrule
    \end{tabular}
    \addtolength{\tabcolsep}{5pt}

    % \vspace{-0.5cm}
    \caption{Accuracy of different models on pristine data, and the attack success rate of adversarial examples generated against different models by AdvGAN on MNIST and CIFAR-10. p: pristine test data; w: \white attack; b-D: black-box attack with dynamic distillation strategy; b-S: black-box attack with static distillation strategy.} 
    \label{t-performance} 
    \vspace{-0.2cm}
    \end{small}
\end{table}

\vspace{-0.4cm}
\paragraph{Models Used in the Experiments}
For MNIST we generate adversarial examples for three models, where 
models A and B are used in \citeauthor{tramer2017ensemble}\ (\citeyear{tramer2017ensemble}).
Model C is the target network architecture used in \citeauthor{carlini2017towards} (\citeyear{carlini2017towards}). For CIFAR-10, we select ResNet-32 and Wide ResNet-34 \cite{he2016deep,zagoruyko2016wide}. Specifically, we use a 32-layer ResNet implemented in TensorFlow\footnote{github.com/tensorflow/models/blob/master/research/ResNet} and Wide ResNet derived from the variant of ``w32-10 wide.''\footnote{github.com/MadryLab/cifar10\_challenge/blob/master/model.py}
We show the classification accuracy of pristine MNIST and CIFAR-10 test data (p) and attack success rate of adversarial examples generated by AdvGAN on different models in Table~\ref{t-performance}. 
\vspace{-0.2cm}
\subsection{AdvGAN in \white Setting}
We evaluate AdvGAN on $f$ with different architectures for MNIST and CIFAR-10.
We first apply AdvGAN to perform \white attack against different models on MNIST dataset.
From the performance of \white attack (Attack Rate (w)) in Table~\ref{t-performance}, we can see that AdvGAN is able to generate adversarial instances to attack all models with high attack success rate.

\begin{figure}[t]
\centering
\begin{minipage}{.15\textwidth}
 \begin{subfigure}{\textwidth}
%  \centerline{Target class}
 \begin{small}\hspace{1pt}0 1 2 3 4 5 6 7 8 9\end{small} \hspace{1pt}\vspace{2pt} \linebreak
 \includegraphics[width=\textwidth]{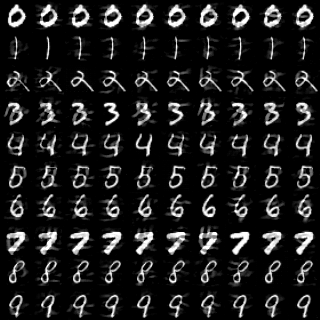}
%  \caption{Model A}
 \label{fig:mnist-wb-one-A}
 \end{subfigure}
\end{minipage}
\begin{minipage}{.15\textwidth}
 \begin{subfigure}{\textwidth}
%  \centerline{Target class}
 \hspace{1pt}\begin{small}0 1 2 3 4 5 6 7 8 9\end{small}\hspace{1pt}\vspace{2pt} \linebreak
 \includegraphics[width=\textwidth]{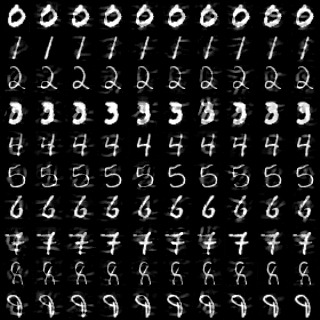}
%  \caption{Model B}
 \label{fig:mnist-wb-one-B}
 \end{subfigure}
\end{minipage}
\begin{minipage}{.15\textwidth}
 \begin{subfigure}{\textwidth}
%  \centerline{Target class}
 \hspace{1pt}\begin{small}0 1 2 3 4 5 6 7 8 9\end{small}\hspace{1pt}\vspace{2pt} \linebreak
 \includegraphics[width=\textwidth]{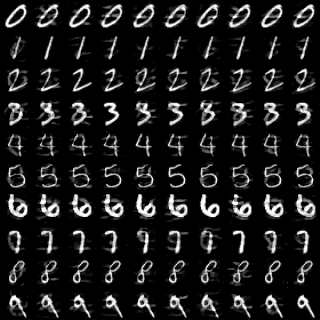}
%  \caption{Model C}
 \label{fig:mnist-wb-one-C}
 \end{subfigure}
\end{minipage}

\begin{minipage}{.15\textwidth}
 \begin{subfigure}{\textwidth}
 \centering
 \includegraphics[width=\textwidth]{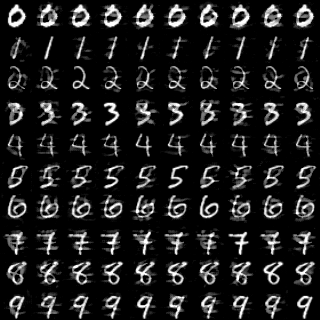}
%  \caption{Model A}
 \label{fig:mnist-bb-one-A}
 \end{subfigure}
\end{minipage}
\begin{minipage}{.15\textwidth}
 \begin{subfigure}{\textwidth}
 \centering
 \includegraphics[width=\textwidth]{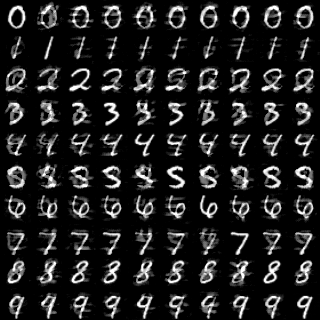}
%  \caption{Model B}
 \label{fig:mnist-bb-one-B}
 \end{subfigure}
\end{minipage}
\begin{minipage}{.15\textwidth}
 \begin{subfigure}{\textwidth}
 \centering
 \includegraphics[width=\textwidth]{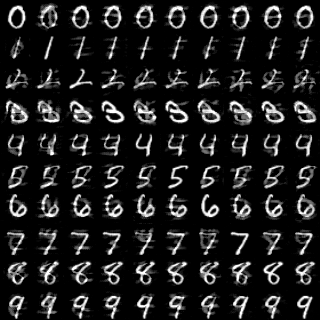}
%  \caption{Model C}
 \label{fig:mnist-bb-one-C}
 \end{subfigure}
\end{minipage}
\vspace{-0.3cm}
\caption{Adversarial examples generated from the same original image to different targets by AdvGAN on MNIST. Row 1: \white attack; Row 2: black-box attack. Left to right: models A, B, and C.On the diagonal, the original images are shown, and the numer on the top denote the targets.\vspace{-0.5cm}} 
\label{fig:mnist-white}
\end{figure}

We also generate adversarial examples from the same original instance $x$, targeting other different classes, as shown in Figures~\ref{fig:mnist-white}.
In the \white setting on MNIST (a)-(c), we can see that the generated adversarial examples for different models appear close to the ground truth/pristine images (lying on the diagonal of the matrix). 

In addition, we analyze the attack success rate based on different loss functions on MNIST. Under the same bounded perturbations (0.3),
if we replace the full loss function in~\eqref{eq:full_loss} with $\mathcal{L}=||\mathcal{G}(x)||_2+\mathcal{L}^f_{\text{adv}}$, which is similar to the objective used in
\citeauthor{baluja2017adversarial},  
the attack success rate becomes 86.2\%. If we replace the loss function with $\mathcal{L}=\mathcal{L}_{\text{hinge}}+\mathcal{L}^f_{\text{adv}}$, the attack success rate is 91.1\%, compared to that of AdvGAN, 98.3\%.

\begin{figure}[t]
\centering
\begin{minipage}{.23\textwidth}
 \begin{subfigure}{\textwidth}
 \centering
 \includegraphics[width=\textwidth]{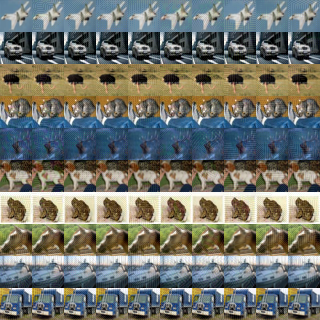}
 \caption{Semi-whitebox attack}
 \label{fig:cifar10-wb-one}
 \end{subfigure}
\end{minipage}
\begin{minipage}{.23\textwidth}
 \begin{subfigure}{\textwidth}
 \centering
 \includegraphics[width=\textwidth]{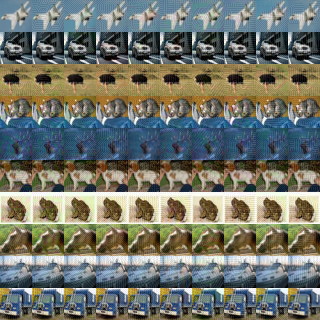}
 \caption{Black-box attack}
 \label{fig:cifar10-bb-one}
 \end{subfigure}
\end{minipage}

\vspace{-0.1cm}
\caption{Adversarial examples generated by AdvGAN on CIFAR-10 for (a) \white attack and (b) black-box attack. Image from each class is perturbed to other different classes. On the diagonal, the original images are shown.} 
\vspace{-0.3cm}
\label{fig:cifar10-same}
\end{figure}

Similarly, on CIFAR-10, we apply the same \white attack for ResNet and Wide ResNet based on AdvGAN,
and  Figure~\ref{fig:cifar10-same} (a) shows some adversarial examples, which are perceptually realistic.

We show adversarial examples for the same original instance targeting different other classes. It is clear that with different targets, the adversarial examples keep similar visual quality compared to the pristine instances on the diagonal. 

\vspace{-0.2cm}
\subsection{AdvGAN in Black-box Setting}
% In this section, we evaluate the performance of AdvGAN for the black-box attack. 
Our black-box attack here is based on dynamic distillation strategy.
We construct a local model to distill model $f$, and we select the architecture of Model C as our local model.
Note that we randomly select a subset of instances disjoint from the training data of AdvGAN to train the local model; that is, we assume the adversaries do not have any prior knowledge of the training data or the model itself.
With the dynamic distillation strategy, the adversarial examples generated by AdvGAN achieve an attack success rate, above $90\%$ for MNIST and $80\%$ for CIFAR-10, compared to $30\%$ and $10\%$ with the static distillation approach, as shown in Table~\ref{t-performance}.

We apply AdvGAN to generate adversarial examples for the same instance targeting different classes on MNIST and randomly select some instances to show in Figure~\ref{fig:mnist-white} (d)-(f). By comparing with the pristine instances on the diagonal, we can see that these adversarial instances can achieve high perceptual quality as the original digits. Specifically, the original digit is somewhat highlighted by adversarial perturbations, which implies a type of perceptually realistic manipulation.
Figure~\ref{fig:cifar10-same}(b) shows similar results for adversarial examples generated on CIFAR-10. These adversarial instances appear photo-realistic compared with the original ones on the diagonal.
\vspace{-0.2cm}
\subsection{Attack Effectiveness Under Defenses}
Facing different types of attack strategies, various defenses have been provided. Among them, different types of adversarial training methods are the most effective. Other categories of defenses, such as those which pre-process an input have mostly been defeated by adaptive attacks \cite{he2017adversarial,carlini2017adversarial}. \citeauthor{goodfellow2014explaining} first propose adversarial training as an effective way to improve the robustness of DNNs, and \citeauthor{tramer2017ensemble} extend it to ensemble adversarial learning.  \citeauthor{madry_towards_2017} have also proposed robust networks against adversarial examples based on well-defined adversaries.
Given the fact that AdvGAN strives to generate adversarial instances from the underlying true data distribution, it can essentially produce more photo-realistic adversarial perturbations compared with other attack strategies. Thus, AdvGAN could have a higher chance to produce adversarial examples that are resilient under different defense methods. In this section, we quantitatively evaluate this property for AdvGAN compared with other attack strategies. 
\vspace{-0.4cm}
\paragraph{Threat Model} As shown in the literature, most of the current defense strategies are not robust when attacking against them \cite{carlini2017towards,he2017adversarial}. Here we consider a weaker threat model, where the adversary is not aware of the defenses and directly tries to attack the original learning model, which is also the first threat model analyzed in \citeauthor{carlini2017towards}. 
In this case, if an adversary can still successfully attack the model, it implies the robustness of the attack strategy. 
Under this setting, we first apply different attack methods to generate adversarial examples based on the original model without being aware of any defense.  
Then we apply different defenses to directly defend against these adversarial instances. 
\vspace{-0.4cm}
\paragraph{Semi-whitebox Attack} 
First, we consider the \white attack setting, where the adversary has white-box access to the model architecture as well as the parameters. 
Here, we replace $f$ in \reffig{framework} with our model A, B, and C, respectively. As a result, adversarial examples will be generated against different models.
We use three adversarial training defenses to train different models for each model architecture:
standard FGSM adversarial training (Adv.) \cite{goodfellow2014explaining},
ensemble adversarial training (Ens.) \cite{tramer2017ensemble},\footnote{
Each ensemble adversarially trained model is trained using (i) pristine training data, (ii) FGSM adversarial examples generated for the current model under training, and (iii) FGSM adversarial examples generated for naturally trained models of two architectures different from the model under training.} and
iterative training (Iter. Adv.) \cite{madry_towards_2017}.
We evaluate the effectiveness of these attacks against these defended models.
In Table~\ref{t-defense-a}, we show that the attack success rate of adversarial examples generated by AdvGAN on different models is higher than those of FGSM and Opt. \cite{carlini2017towards}.

\begin{table}[h]
\begin{small}
    \centering
    \begin{tabular}{ c| c | c c c c}
        \toprule
     Data & Model  & Defense & FGSM & Opt.  & {\bf AdvGAN} \\ \hline 
        \hline
    \multirow{9}{*}{\shortstack{M\\N\\I\\S\\T}} &  \multirow{3}{*}{A} &  Adv. & 4.3\% & 4.6\%  & {\bf 8.0\%} \\
                    &     & Ens.       & 1.6\% & 4.2\%  &{\bf 6.3\%} \\
                     &    & Iter.Adv.  & 4.4\% & 2.96\% &{\bf 5.6\%} \\
        % \hline
        \cline{2-6}
        & \multirow{3}{*}{B} & Adv.            & 6.0\% & 4.5\%  & {\bf7.2\% }\\
          &               & Ens.       & 2.7\% & 3.18\% &{\bf 5.8\%} \\
           &              & Iter.Adv.  &{\bf 9.0}\% & 3.0\% & 6.6\% \\
        \cline{2-6}
        & \multirow{3}{*}{C} & Adv.            & 2.7\% & 2.95\% & {\bf18.7\%} \\
          &               & Ens.       & 1.6\% & 2.2\%  & {\bf13.5\%} \\
           &              & Iter.Adv.  & 1.6\% & 1.9\%  &{\bf 12.6\%} \\
        \cline{1-6}
        \multirow{6}{*}{\shortstack{C\\I\\F\\A\\R\\10}}&   \multirow{3}{*}{ResNet} & Adv.  & 13.10\% & 11.9\%  & {\bf 16.03\%} \\
         &                & Ens.   & 10.00\% & 10.3\%  & {\bf14.32\%}\\
         &                & Iter.Adv  & 22.8\% & 21.4\% &{\bf 29.47\% }\\
        \cline{2-6}
      &  \multirow{3}{*}{\shortstack{Wide\\ResNet}} & Adv.            & 5.04\% & 7.61\%  & {\bf 14.26\%} \\
        &                 & Ens.       & 4.65\% & 8.43\% & {\bf 13.94 \%} \\
         &                & Iter.Adv.  & 14.9\% & 13.90\% & {\bf 20.75\%} \\
  
        \bottomrule
        
    \end{tabular}

    \caption{Attack success rate of adversarial examples generated by AdvGAN in \white setting, and other white-box attacks under defenses on MNIST and CIFAR-10. } 
    \label{t-defense-a}
    \vspace{-0.2cm}
    
    \end{small}
\end{table}

\begin{table}[htb]
\begin{small}
    \centering
    
    \addtolength{\tabcolsep}{-4pt}
    \begin{tabular}{c|ccc| ccc}
        \toprule
      & \multicolumn{3}{ c |}{MNIST} &  \multicolumn{3}{ c }{CIFAR-10}   \\      
        Defense &FGSM & Opt.  & {\bf AdvGAN} &FGSM & Opt.  & {\bf AdvGAN}  \\ \hline 
        \hline
      Adv.  &  3.1\% & 3.5\%   & {\bf11.5\% }& 13.58\%  & 10.8\% & {\bf 15.96\%} \\
      Ens.   &2.5\%   &3.4\% & {\bf 10.3\% } & 10.49\%   &9.6\% & {\bf 12.47\%}\\
      Iter.Adv.  &2.4\% & 2.5\%    & {\bf 12.2\%} & 22.96\% & 21.70\% &{\bf 24.28\%}  \\
        \bottomrule
    \end{tabular}
    \addtolength{\tabcolsep}{4pt}
    % \vspace{-0.2cm}
    \caption{Attack success rate of adversarial examples generated by different black-box adversarial strategies under defenses on MNIST and CIFAR-10} 
        \label{t-defense-c} 
    \vspace{-0.2cm}

\end{small}
\end{table}

\begin{table}[t!]
\begin{small}
    \centering
     \begin{tabular}{lcc}
        \toprule
        Method & Accuracy (xent loss) & Accuracy (cw loss) \\ \hline 
        \hline
        FGSM  & 95.23\% & 96.29\% \\ 
        % FGSM (cw loss)  & 96.29\% \\ 
        PGD & 93.66\% & 93.79\%\\
        % PGD & 93.79\%\\
        Opt  & - & 91.69\% \\
        \textbf{AdvGAN} & - & \textbf{88.93\%} \\
        \bottomrule
    \end{tabular}
    % \vspace{-0.5cm}
    \caption{Accuracy of the MadryLab public model under different attacks in white-box setting. AdvGAN here achieved the best performance.} 
     \vspace{-0.2cm}
    \label{t-madry} 
    
    \end{small}
\end{table}

\noindent\textbf{Black-box Attack}
For AdvGAN, we use model B as the black-box model and train a distilled model to perform black-box attack against model B and report the attack success rate in Table~\ref{t-defense-c}.
For the black-box attack comparison purpose, transferability based attack is applied for FGSM and Opt.
We use FGSM and Opt. to attack model A on MNIST, and we use these adversarial examples to test on model B and report the corresponding classification accuracy. 
We can see that the adversarial examples generated by the black-box AdvGAN consistently achieve much higher attack success rate compared with other attack methods. 

For CIFAR-10, we use a ResNet as the black-box model and train a distilled model to perform black-box attack against the ResNet. To evaluate black-box attack for optimization method and FGSM, we use adversarial examples generated by attacking Wide ResNet and test them on ResNet to report black-box attack results for these two methods.

In addition, we apply AdvGAN to the MNIST challenge. Among all the standard attacks shown in Table~\ref{t-madry}, AdvGAN achieve 88.93\% in the white-box setting. 

Among reported black-box attacks, AdvGAN achieved an accuracy of 92.76\%, outperforming all other state-of-the-art attack strategies submitted to the challenge.

\begin{figure}[tbh]
\centering
 \begin{subfigure}{.23\textwidth}
 \centering
 \includegraphics[width=\textwidth]{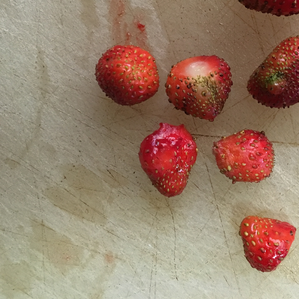}
 \caption{Strawberry}
 \label{fig:in_ori}
 \end{subfigure}
 \begin{subfigure}{.23\textwidth}
 \centering
 \includegraphics[width=\textwidth]{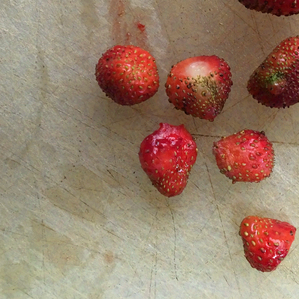}
 \caption{ Toy poodle}
 \label{fig:in_fgsm}
 \end{subfigure}
 \begin{subfigure}{.23\textwidth}
 \centering
 \includegraphics[width=\textwidth]{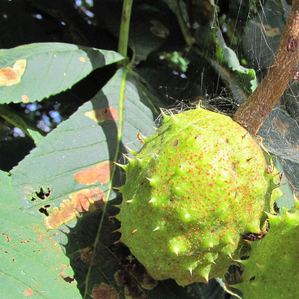}
 \caption{Buckeye}
 \label{fig:in_cw}
 \end{subfigure}
 \begin{subfigure}{.23\textwidth}
 \centering
 \includegraphics[width=\textwidth]{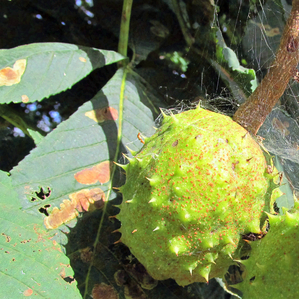}
 \caption{Toy poodle}
 \label{fig:in_stn}
 \end{subfigure}
\vspace{-0.3cm}
\caption{Examples from an ImageNet-compatible set, and the labels denote corresponding classification results
Left: original benign images; right: adversarial images generated by AdvGAN against Inception\_v3. \vspace{-0.4cm} }
\label{fig:extra-imagenet}
\end{figure}

\subsection{High Resolution Adversarial Examples }
To evaluate AdvGAN's ability of generating high resolution adversarial examples, we attack against Inception\_v3 and quantify attack success rate and perceptual realism of generated adversarial examples.

\noindent\textbf{Experiment settings.} In the following experiments, we select 100 benign images from the DEV dataset of the NIPS 2017 adversarial attack competition \cite{imagenet_challenge}.
This competition provided a dataset compatible with ImageNet.
We generate adversarial examples (299$\times$299 pixels), each targeting a random incorrect class, with
$L_\infty$ bounded within 0.01 for Inception\_v3. The attack success rate is 100\%. 

In Figure~\ref{fig:extra-imagenet}, we show some randomly selected examples of original and adversarial examples generated by AdvGAN.
\vspace{-0.5cm}
\paragraph{Human Perceptual Study.}
We validate the realism of AdvGAN's adversarial examples with a user study on Amazon Mechanical Turk (AMT).
We use 100 pairs of original images and adversarial examples (generated as described above) and ask workers to choose which image of a pair is more \textit{visually realistic}.

Our study follows a protocol from \citeauthor{isola2017image}, where a worker is shown a pair of images for 2 seconds, then the worker has unlimited time to decide.
We limit each worker to at most $20$ of these tasks.
We collected 500 choices, about 5 per pair of images, from 50 workers on AMT.
The AdvGAN examples were chosen as more realistic than the original image in $49.4\% \pm 1.96\%$ of the tasks (random guessing would result in about $50\%$).
This result show that these high-resolution AdvGAN adversarial examples are about as realistic as benign images.
% \vspace{-0.3cm}
\section{Conclusion} %and Future Work}
In this paper, we propose AdvGAN to generate adversarial examples using generative adversarial networks (GANs).
% , which have been applied to different domains including images, text, and speech generation. 
In our AdvGAN framework, once trained, the feed-forward generator can produce adversarial perturbations efficiently. It can also perform both \white and black-box attacks with high attack success rate. 
In addition, when we apply AdvGAN to generate adversarial instances on different models without knowledge of the defenses in place, the generated adversarial examples can preserve high perceptual quality and attack the state-of-the-art defenses with higher attack success rate than examples generated by the competing methods. 
This property makes AdvGAN a promising candidate for improving adversarial training defense methods.

\section*{Acknowledgments}
We thank Weiwei Hu for his valuable discussions
on this work. This work was supported in part by Berkeley Deep Drive, JD.COM, the Center for Long-Term Cybersecurity, and FORCES (Foundations Of Resilient CybEr-Physical Systems), which receives support from the National Science Foundation (NSF award numbers CNS-1238959, CNS-1238962, CNS-1239054, CNS-1239166, CNS-1422211 and CNS-1616575).
% \newpage
\bibliographystyle{named}
\bibliography{iclr2018_conference}

\end{document}

%% file: def.tex
\newcommand{\white}[0]{{semi-whitebox}\xspace}
\newcommand{\White}[0]{{Semi-whitebox}\xspace}

%% file: main.bbl
\begin{thebibliography}{}

\bibitem[\protect\citeauthoryear{Baluja and
  Fischer}{2017}]{baluja2017adversarial}
Shumeet Baluja and Ian Fischer.
\newblock Adversarial transformation networks: Learning to generate adversarial
  examples.
\newblock {\em arXiv preprint arXiv:1703.09387}, 2017.

\bibitem[\protect\citeauthoryear{Carlini and
  Wagner}{2017a}]{carlini2017adversarial}
Nicholas Carlini and David Wagner.
\newblock Adversarial examples are not easily detected: Bypassing ten detection
  methods.
\newblock In {\em Proceedings of the 10th ACM Workshop on Artificial
  Intelligence and Security}, pages 3--14. ACM, 2017.

\bibitem[\protect\citeauthoryear{Carlini and
  Wagner}{2017b}]{carlini2017towards}
Nicholas Carlini and David Wagner.
\newblock Towards evaluating the robustness of neural networks.
\newblock In {\em Security and Privacy (SP), 2017 IEEE Symposium on}, pages
  39--57. IEEE, 2017.

\bibitem[\protect\citeauthoryear{Deng \bgroup \em et al.\egroup
  }{2009}]{deng2009imagenet}
Jia Deng, Wei Dong, Richard Socher, Li-Jia Li, Kai Li, and Li~Fei-Fei.
\newblock Imagenet: A large-scale hierarchical image database.
\newblock In {\em CVPR}, pages 248--255. IEEE, 2009.

\bibitem[\protect\citeauthoryear{Evtimov \bgroup \em et al.\egroup
  }{2017}]{evtimov2017robust}
Ivan Evtimov, Kevin Eykholt, Earlence Fernandes, Tadayoshi Kohno, Bo~Li, Atul
  Prakash, Amir Rahmati, and Dawn Song.
\newblock Robust physical-world attacks on machine learning models.
\newblock {\em arXiv preprint arXiv:1707.08945}, 2017.

\bibitem[\protect\citeauthoryear{Goodfellow \bgroup \em et al.\egroup
  }{2014}]{goodfellow2014generative}
Ian Goodfellow, Jean Pouget-Abadie, Mehdi Mirza, Bing Xu, David Warde-Farley,
  Sherjil Ozair, Aaron Courville, and Yoshua Bengio.
\newblock Generative adversarial nets.
\newblock In {\em NIPS}, pages 2672--2680, 2014.

\bibitem[\protect\citeauthoryear{Goodfellow \bgroup \em et al.\egroup
  }{2015}]{goodfellow2014explaining}
Ian Goodfellow, Jonathon Shlens, and Christian Szegedy.
\newblock Explaining and harnessing adversarial examples.
\newblock In {\em International Conference on Learning Representations}, 2015.

\bibitem[\protect\citeauthoryear{He \bgroup \em et al.\egroup
  }{2016}]{he2016deep}
Kaiming He, Xiangyu Zhang, Shaoqing Ren, and Jian Sun.
\newblock Deep residual learning for image recognition.
\newblock In {\em CVPR}, pages 770--778, 2016.

\bibitem[\protect\citeauthoryear{He \bgroup \em et al.\egroup
  }{2017}]{he2017adversarial}
Warren He, James Wei, Xinyun Chen, Nicholas Carlini, and Dawn Song.
\newblock Adversarial example defenses: Ensembles of weak defenses are not
  strong.
\newblock {\em arXiv preprint arXiv:1706.04701}, 2017.

\bibitem[\protect\citeauthoryear{Hinton \bgroup \em et al.\egroup
  }{2015}]{hinton2015distilling}
Geoffrey Hinton, Oriol Vinyals, and Jeff Dean.
\newblock Distilling the knowledge in a neural network.
\newblock {\em arXiv preprint arXiv:1503.02531}, 2015.

\bibitem[\protect\citeauthoryear{Hu and Tan}{2017}]{hu2017generating}
Weiwei Hu and Ying Tan.
\newblock Generating adversarial malware examples for black-box attacks based
  on {GAN}.
\newblock {\em arXiv preprint arXiv:1702.05983}, 2017.

\bibitem[\protect\citeauthoryear{Isola \bgroup \em et al.\egroup
  }{2017}]{isola2017image}
Phillip Isola, Jun-Yan Zhu, Tinghui Zhou, and Alexei~A Efros.
\newblock Image-to-image translation with conditional adversarial networks.
\newblock {\em CVPR}, 2017.

\bibitem[\protect\citeauthoryear{Krizhevsky and
  Hinton}{2009}]{krizhevsky2014cifar}
Alex Krizhevsky and Geoffrey Hinton.
\newblock Learning multiple layers of features from tiny images.
\newblock 2009.

\bibitem[\protect\citeauthoryear{Kurakin \bgroup \em et al.\egroup
  }{2018}]{imagenet_challenge}
Alexey Kurakin, Ian Goodfellow, Samy Bengio, Yinpeng Dong, Fangzhou Liao, Ming
  Liang, Tianyu Pang, Jun Zhu, Xiaolin Hu, Cihang Xie, et~al.
\newblock Adversarial attacks and defences competition.
\newblock {\em arXiv preprint arXiv:1804.00097}, 2018.

\bibitem[\protect\citeauthoryear{LeCun and Cortes}{1998}]{lecun1998mnist}
Yann LeCun and Corrina Cortes.
\newblock The {MNIST} database of handwritten digits.
\newblock 1998.

\bibitem[\protect\citeauthoryear{Liu \bgroup \em et al.\egroup
  }{2017}]{liu2016delving}
Yanpei Liu, Xinyun Chen, Chang Liu, and Dawn Song.
\newblock Delving into transferable adversarial examples and black-box attacks.
\newblock In {\em ICLR}, 2017.

\bibitem[\protect\citeauthoryear{Mao \bgroup \em et al.\egroup
  }{2017}]{mao2016least}
Xudong Mao, Qing Li, Haoran Xie, Raymond~YK Lau, Zhen Wang, and Stephen~Paul
  Smolley.
\newblock Least squares generative adversarial networks.
\newblock In {\em 2017 IEEE International Conference on Computer Vision
  (ICCV)}, pages 2813--2821. IEEE, 2017.

\bibitem[\protect\citeauthoryear{Mądry \bgroup \em et al.\egroup
  }{2017}]{madry_towards_2017}
Aleksander Mądry, Aleksandar Makelov, Ludwig Schmidt, Dimitris Tsipras, and
  Adrian Vladu.
\newblock Towards deep learning models resistant to adversarial attacks.
\newblock {\em arXiv:1706.06083 [cs, stat]}, June 2017.

\bibitem[\protect\citeauthoryear{Papernot \bgroup \em et al.\egroup
  }{2016}]{papernot2016practical}
Nicolas Papernot, Patrick McDaniel, Ian Goodfellow, Somesh Jha, Z~Berkay Celik,
  and Ananthram Swami.
\newblock Practical black-box attacks against deep learning systems using
  adversarial examples.
\newblock {\em arXiv preprint}, 2016.

\bibitem[\protect\citeauthoryear{Szegedy \bgroup \em et al.\egroup
  }{2014}]{szegedy2014intriguing}
Christian Szegedy, Wojciech Zaremba, Ilya Sutskever, Joan Bruna, Dumitru Erhan,
  Ian Goodfellow, and Rob Fergus.
\newblock Intriguing properties of neural networks.
\newblock In {\em ICLR}, 2014.

\bibitem[\protect\citeauthoryear{Tram{\`e}r \bgroup \em et al.\egroup
  }{2017}]{tramer2017ensemble}
Florian Tram{\`e}r, Alexey Kurakin, Nicolas Papernot, Dan Boneh, and Patrick
  McDaniel.
\newblock Ensemble adversarial training: Attacks and defenses.
\newblock {\em arXiv preprint arXiv:1705.07204}, 2017.

\bibitem[\protect\citeauthoryear{Xiao \bgroup \em et al.\egroup
  }{2018}]{xiao2018spatially}
Chaowei Xiao, Jun-Yan Zhu, Bo~Li, Warren He, Mingyan Liu, and Dawn Song.
\newblock Spatially transformed adversarial examples.
\newblock {\em arXiv preprint arXiv:1801.02612}, 2018.

\bibitem[\protect\citeauthoryear{Zagoruyko and
  Komodakis}{2016}]{zagoruyko2016wide}
Sergey Zagoruyko and Nikos Komodakis.
\newblock Wide residual networks.
\newblock {\em arXiv preprint arXiv:1605.07146}, 2016.

\bibitem[\protect\citeauthoryear{Zhu \bgroup \em et al.\egroup
  }{2016}]{zhu2016generative}
Jun-Yan Zhu, Philipp Kr{\"a}henb{\"u}hl, Eli Shechtman, and Alexei~A Efros.
\newblock Generative visual manipulation on the natural image manifold.
\newblock In {\em ECCV}, pages 597--613. Springer, 2016.

\bibitem[\protect\citeauthoryear{Zhu \bgroup \em et al.\egroup
  }{2017}]{zhu2017unpaired}
Jun-Yan Zhu, Taesung Park, Phillip Isola, and Alexei~A Efros.
\newblock Unpaired image-to-image translation using cycle-consistent
  adversarial networks.
\newblock {\em ICCV}, pages 2242--2251, 2017.

\end{thebibliography}
